\documentclass[12pt]{article}
\usepackage{epsfig,amsmath}

\title{\bf Relativistic Static Thin Disks: \\ The Counter-Rotating Model}

\author{Omar A. Espitia\thanks{e-mail: oespitia@uis.edu.co} \ \ and \
Guillermo A. Gonz\'{a}lez\thanks{e-mail: guillego@uis.edu.co}	\\
{\it Escuela de F\'{\i}sica, Universidad Industrial de Santander}	\\
{\it A.A. 678, Bucaramanga, Colombia}}

\begin{document}

\maketitle

\begin{abstract}

A detailed study of the Counter-Rotating Model (CRM) for generic
finite static axially symmetric thin disks with nonzero radial pressure
is presented. We find a general constraint over the counter-rotating
tangential velocities needed to cast the surface energy-momentum tensor of
the disk as the superposition of two counter-rotating perfect fluids. We
also found expressions for the energy density and pressure of the
counter-rotating fluids. Then we shown that, in general, there is not
possible to take the two counter-rotating fluids as circulating along
geodesics neither take the two counter-rotating tangential velocities
as equal and opposite. An specific example is studied where we obtain
some CRM with well defined counter-rotating tangential velocities and
stable against radial perturbations. The CRM obtained are in agree with
the strong energy condition, but there are regions of the disks with
negative energy density, in violation of the weak energy condition.
\vspace{0.4cm}

\end{abstract}

\newpage
\section{Introduction}

The study of axially symmetric solutions of Einstein field equations
corresponding to disklike configurations of matter has a long history.
These were first studied by Bonnor and Sackfield \cite{BS}, obtaining
pressureless static disks, and by Morgan and Morgan, obtaining static
disks with and without radial pressure \cite{MM1,MM2}. In connection with
gravitational collapse, disks were first studied by Chamorro, Gregory and
Stewart \cite{CHGS}. In the last years, disks models with radial tension
\cite{GL1}, magnetic fields \cite{LET1} and magnetic and electric fields
\cite{KBL} have been also studied. Several classes of exact solutions
of the Einstein field equations corresponding to static and stationary
thin disks have been obtained by different authors [\ref{bib:LP} --
\ref{bib:GL2}], with or without radial pressure.

In the case of static disks without radial pressure, there are two common
interpretations. The stability of these models can be explained by either
assuming the existence of hoop stresses or that the particles on the disk
plane move under the action of their own gravitational field in such a
way that as many particles move clockwise as counterclockwise. This last
interpretation, the ``Counter-Rotating Model'' (CRM), is frequently made
since it can be invoked to mimic true rotational effects. Even though
this interpretation can be seen as a device, there are observational
evidence of disks made of streams of rotating and counter-rotating
matter \cite{RGK,RFF}.

Usually has been considered that the CRM can be applied only when
we do not have radial pressure and the azimuthal stress is positive
(pressure). These conditions, however, are very restrictive and, in
many cases, we have disks models that only agree with them in a partial
region. Thus, the CRM will be valid only as a partial interpretation
of the corresponding disks. Another, common, assumption is to take the
CRM as representing two fluids that circulate in opposite directions
with the same tangential velocity. As we will show in this paper, this
is not the case and, in general, the two fluids circulate with different
velocities. Furthermore, in some cases may not be possible to obtain a CRM
if the two tangential velocities are taken as equal and opposite. Also,
commonly is assumed that the two counter-rotating fluids must be taken as
circulating along geodesics. We also will show that this is not necessary
and that only can be made if the radial pressure is constant.

The aim of this paper is a detailed study of the CRM for generic finite
static axially symmetric thin disks with nonzero radial pressure. In
Sec. 2 we present a summary of the procedure to obtain these thin disks
models and obtain the surface energy-momentum tensor of the disk. In
the next section, Sec. 3, we consider the CRM for the disk. We find
a general constraint over the counter-rotating tangential velocities
needed to cast the surface energy-momentum tensor of the disk as the
superposition of two counter-rotating perfect fluids. We also found
expressions for the energy density and pressure of the counter-rotating
fluids. Then we shown that, in general, there is not possible to take the
two counter-rotating tangential velocities as equal and opposite neither
take the two counter-rotating fluids as circulating along geodesics. In
Sec. 4, we consider an specific example where we obtain some CRM with
well defined counter-rotating tangential velocities and stable against
radial perturbations. The CRM obtained are in agree with the strong
energy condition, but there are regions of the disks with negative energy
density, in violation of the weak energy condition. Finally, in Sec. 5,
we summarize our main results.

\section{Relativistic Static Thin Disks}

In this section we present, following closely reference \cite{GL1},
a summary of the procedure to obtain finite static axially symmetric
thin disks with nonzero radial pressure. The metric can be written as
the line element,
\begin{equation}
ds^2 = e^{- 2 \Phi} [{\cal R}^2 d\varphi^2 + e^{2 \Lambda} (dr^2 + dz^2)]
\ - \ e^{2 \Phi} dt^2 \ , \label{eq:met}
\end{equation}
where $\Phi$, $\Lambda$ and ${\cal R}$ are functions of $r$ and $z$
only. The Einstein vacuum equations for this metric are equivalent to
the system
\begin{subequations}\begin{eqnarray}
&	&{\cal R}_{,rr} \ + \ {\cal R}_{,zz} \ = \ 0 \ ,\label{eq:eins1}\\
&	&	\nonumber						\\
&	&( {\cal R} \Phi_{,r} )_{,r} + ( {\cal R} \Phi_{,z} )_{,z}
\ = \ 0 \ , \label{eq:eins2}						\\
&	&	\nonumber						\\
&	&{\cal R}_{,z} \Lambda_{,r} + {\cal R}_{,r} \Lambda_{,z} - 2 {\cal
R} \Phi_{,r} \Phi_{,z} - {\cal R}_{,rz} \ = \ 0 \ , \label{eq:eins3}	\\
&	&		\nonumber					\\
&	&{\cal R}_{,r} \Lambda_{,r} - {\cal R}_{,z} \Lambda_{,z} -
{\cal R} ( {\Phi_{,r}}^2 - {\Phi_{,z}}^2 ) + {\cal R}_{,zz} \ = \ 0 \
, \label{eq:eins4}
\end{eqnarray}\label{eq:eins}\end{subequations}
where we assume the existence of the second derivatives of the functions
$\Phi$, $\Lambda$ and ${\cal R}$.

A general solution of the above system can be obtained by the following
procedure. Let $\nu = r + i z$ and ${\cal F} (\nu)$ any analytical
function of $\nu$. Then we take
\begin{subequations}
\begin{eqnarray}
{\cal R} (r,z) \ &=& \ {\rm Re} \ {\cal F} (\nu) \ , \label{eq:com1}\\
&&	\nonumber	\\
{\cal Z} (r,z) \ &=& \ {\rm Im} \ {\cal F} (\nu) \ , \label{eq:com2}\\
&&	\nonumber	\\
\Phi (r,z) \ &=& \ \Psi ({\cal R},{\cal Z}) \ , \label{eq:com3}\\
&&	\nonumber	\\
\Lambda (r,z) \ &=& \ \Pi ({\cal R},{\cal Z}) \ + \ \ln
|{\cal F}' (\nu)| \ , \label{eq:com4}
\end{eqnarray}\label{eq:com}
\end{subequations}
where $\Psi ({\cal R},{\cal Z})$ and $\Pi ({\cal R},{\cal Z})$ are
solutions of the Weyl equations \cite{WEY1,WEY2}
\begin{subequations}
\begin{eqnarray}
&	&( {\cal R} \Psi_{,{\cal R}} )_{,{\cal R}} + ( {\cal R}
\Psi_{,{\cal Z}} )_{,{\cal Z}} \ = \ 0 \ , \label{eq:weyl1}		\\
&	&	\nonumber						\\
&	&\Pi_{,{\cal R}} \ = \ {\cal R} ( {\Psi_{,{\cal R}}}^2 -
{\Psi_{,{\cal Z}}}^2 ) \ ,  \label{eq:weyl2}	    \\
&	&		\nonumber					\\
&	&\Pi_{,{\cal Z}} \ = \ 2 {\cal R} \Psi_{,{\cal R}} \Psi_{,{\cal
Z}}\ . \label{eq:weyl3}
\end{eqnarray}
\end{subequations}
Is easy to see that the condition of integrability of the system
(\ref{eq:weyl2}) -- (\ref{eq:weyl3}) is guaranteed by the equation
(\ref{eq:weyl1}). Also we can see that this equation is equivalent
with the Laplace equation in flat three-di\-men\-sio\-nal space for an
axially symmetric function and so $\Psi$ can be taken as a solution for an
appropriated Newtonian source with axial symmetry. Once a  solution $\Psi$
is known, $\Pi$ is computed from (\ref{eq:weyl2}) -- (\ref{eq:weyl3})
and so we obtain, from (\ref{eq:com1}) - (\ref{eq:com4}), a solution of
the field equations (\ref{eq:eins1}) - (\ref{eq:eins4}).

Now if we assume that ${\cal R}$, $\Phi$ and $\Lambda$ are symmetrical
functions of $z$ and that the first derivatives of the metric tensor
are not continuous on the plane $z = 0$, with discontinuity functions
$$
b_{ab} \ = g_{ab,z}|_{_{z = 0^+}} \ - \ g_{ab,z}|_{_{z = 0^-}} \ =
\ 2 \ g_{ab,z}|_{_{z = 0^+}} \ ,
$$
the Einstein equations yield an energy-momentum tensor $T_a^b \ = \
Q_a^b \ \delta (z)$, where $\delta (z)$ is the usual Dirac function with
support on the disk and
$$
Q^a_b = \frac{1}{2}\{b^{az}\delta^z_b - b^{zz}\delta^a_b + g^{az}b^z_b -
g^{zz}b^a_b + b^c_c (g^{zz}\delta^a_b - g^{az}\delta^z_b)\}
$$
is the distributional energy-momentum tensor. The ``true'' surface
energy-momentum tensor (SEMT) of the disk, $S_a^b$, can be obtained
through the relation
\begin{equation}
S_a^b \ = \ \int T_a^b \ ds_n \ = \ e^{\Lambda -
\Phi} \ Q_a^b \ ,
\end{equation}
where $ds_n = \sqrt{g_{zz}} \ dz$ is the ``physical measure'' of length
in the normal to the disk direction. For the metric (\ref{eq:met})
we obtain
\begin{subequations}\begin{eqnarray}
&S^0_0 &= \ 2 e^{\Phi - \Lambda} \left\{ \Lambda,_z - \ 2 \Phi,_z + \
\frac{{\cal R},_z}{{\cal R}} \right\} , \label{eq:emt1}     		\\
&	&	\nonumber	\\
&S^1_1 &= \ 2 e^{\Phi - \Lambda} \Lambda,_z , \label{eq:emt2} \\
&	&	\nonumber	\\
&S^2_2 &= \ 2 e^{\Phi - \Lambda} \left\{ \frac{{\cal R},_z}{\cal R}
\right\} , \label{eq:emt3}
\end{eqnarray}\end{subequations}
where all the quantities are evaluated at $z = 0^+$.

We can write the metric and the SEMT in the canonical forms
\begin{subequations}\begin{eqnarray}
&	&g_{ab} \ = \ - V_a V_b + W_a W_b + X_a X_b + Y_a Y_b \ ,
\label{eq:metdia}							\\
&	&	\nonumber						\\
&	&S_{ab} \ = \ \sigma V_a V_b + p_\varphi W_a W_b + p_r X_a X_b \
, \label{eq:emtdia}
\end{eqnarray}\end{subequations}
with an orthonormal tetrad ${{\rm e}_{\hat a}}^b = \{ V^b , W^b , X^b ,
Y^b \}$, where
\begin{subequations}\begin{eqnarray}
V^a &=& e^{- \Phi} \ ( 1, 0, 0, 0 ) \ ,	\\
	&	&	\nonumber	\\
W^a &=& \frac{e^\Phi}{\cal R} \ \ ( 0, 1, 0, 0 ) \ ,	\\
	&	&	\nonumber	\\
X^a &=& e^{\Phi - \Lambda} ( 0, 0, 1, 0 ) \ ,	\\
	&	&	\nonumber	\\
Y^a &=& e^{\Phi - \Lambda} ( 0, 0, 1, 0 ) \ .
\end{eqnarray}\label{eq:tetrad}\end{subequations}
The energy density, the azimuthal pressure, and the radial pressure are,
respectively,
\begin{equation}
\sigma \ = \ - S^0_0 \quad , \quad p_\varphi \ = \ S^1_1 \quad ,
\quad p_r \ = \ S^2_2 \ ,
\end{equation}
and
\begin{equation}
\varrho \ = \ \sigma \ + \ p_\varphi \ + \ p_r 
\end{equation}
is the effective Newtonian density.

\section{The Counter-Rotating Model}

We now consider that the SEMT $S^{ab}$ can be written as the superposition
of two counter-rotating perfect fluids that circulate in opposite
directions; that is, based on the two-perfect-fluid model of anisotropic
fluids \cite{LET2}, we assume that $S^{ab}$ can be cast as
\begin{equation}
S^{ab} \ = \ S_+^{ab} \ + \ S_-^{ab} \ , \label{eq:emtsum}
\end{equation}
where $S_+^{ab}$ and $S_-^{ab}$ are, respectively, the SEMT of the
prograd and retrograd counter-rotating fluids.

Let be $U_\pm^a = ( U_\pm^0 , U_\pm^1, 0 , 0 )$ the velocity vectors
of the two counter-rotating fluids. In order to do the decomposition
(\ref{eq:emtsum}) we project the velocity vectors onto the tetrad ${{\rm
e}_{\hat a}}^b$, using the relations \cite{CHAN}
\begin{equation}
U_\pm^{\hat a} \ = \ {{\rm e}^{\hat a}}_b U_\pm^b \qquad , \qquad U_\pm^
a = \ U_\pm^{\hat c} {{\rm e}_{\hat c}}^a  .
\end{equation}
With the tetrad (\ref{eq:tetrad}) we can write
\begin{equation}
U_\pm^a \ = \ \frac{ V^a + {\rm U}_\pm W^a }{\sqrt{1 - {\rm U}_\pm^2}}
, \label{eq:vels}
\end{equation}
and thus
\begin{subequations}\begin{eqnarray}
&V^a &= \ \frac{\sqrt{1 - {\rm U}_-^2} {\rm U}_+ U_-^a - \sqrt{1 -
{\rm U}_+^2} {\rm U}_- U_+^a}{{\rm U}_+ - {\rm U}_-} \ , \label{eq:va} \\
&	&	\nonumber	\\
&W^a &= \ \frac{\sqrt{1 - {\rm U}_+^2} U_+^a - \sqrt{1 - {\rm U}_-^2}
U_-^a}{{\rm U}_+ - {\rm U}_-} \ , \label{eq:wa}
\end{eqnarray}\label{eq:vawa}\end{subequations}
where ${\rm U}_\pm = U_\pm^{\hat 1} / U_\pm^{\hat 0}$ are the tangential
velocities of the fluids with respect to the tetrad.

In terms of the metric $h_{ab} = g_{ab} - Y_a Y_b$ of the hypersurface
$z = 0$, we can write the SEMT as
\begin{equation}
S^{ab} \ = \ ( \sigma + p_r ) V^a V^b \ + \ ( p_\varphi - p_r ) W^a W^b \
+ \ p_r h^{ab} ,
\end{equation}
and so, using (\ref{eq:vawa}), we obtain

\begin{eqnarray}
S^{ab} & = & \frac{ f( {\rm U}_- , {\rm U}_- ) (1 - {\rm U}_+^2) \
U_+^a U_+^b }{({\rm U}_+ - {\rm U}_-)^2}	\nonumber	\\
&	&		\nonumber	\\
& + & \frac{ f( {\rm U}_+ , {\rm U}_+ ) (1 - {\rm U}_-^2) \ U_-^a U_-^b
}{({\rm U}_+ - {\rm U}_-)^2}			\nonumber	\\
&	&		\nonumber	\\
& - & \frac{ f( {\rm U}_+ , {\rm U}_- ) (1 - {\rm U}_+^2)^{\frac{1}{2}}
(1 - {\rm U}_-^2)^{\frac{1}{2}} ( U_+^a U_-^b + U_-^a U_+^b ) }{({\rm U}_+
- {\rm U}_-)^2}		\nonumber	\\
&	&		\nonumber	\\
& + & p_r h^{ab} ,	\nonumber
\end{eqnarray}
where
\begin{equation}
f( {\rm U}_1 , {\rm U}_2 ) \ = \ ( \sigma + p_r ) {\rm U}_1 {\rm U}_2 +
p_\varphi - p_r \ . \label{eq:fuu}
\end{equation}
Thus, in order to cast the SEMT in the form (\ref{eq:emtsum}), the mixed
term must be absent and so the counter-rotating tangential velocities
must be related by
\begin{equation}
f( {\rm U}_+ , {\rm U}_- ) \ = \ 0 \ , \label{eq:liga}
\end{equation}
where we assume that $|{\rm U}_\pm| \neq 1$.

Assuming a given choice for the counter-rotating velocities in agree
with the above relation, we can write the SEMT as (\ref{eq:emtsum}) with
\begin{equation}
S^{ab}_\pm = ( \sigma_\pm + p_\pm ) \ U_\pm^a U_\pm^b \ + \ p_\pm \ h^{ab} ,
\end{equation}
where
\begin{subequations}
\begin{eqnarray}
&\sigma_+ + \ p_+ &= \left[ \frac{ 1 - {\rm U}_+^2 }{{\rm U}_- - {\rm U}_+}
\right] \{ ( \sigma + p_r ) {\rm U}_- \} \ , \\
&	&	\nonumber	\\
&\sigma_- + \ p_- &= \left[ \frac{ 1 - {\rm U}_-^2 }{{\rm U}_+ - {\rm U}_-}
\right] \{ ( \sigma + p_r ) {\rm U}_+ \} \ , \\
&	&	\nonumber	\\
&\sigma_+ + \ \sigma_- &= \ \ \sigma \ + \ p_r \ - \ p_\varphi , \\
&	&	\nonumber	\\
&p_+ + \ p_- &= \ \ p_r .
\end{eqnarray}\label{eq:dpcon}
\end{subequations}
Note that the counter-rotating energy densities $\sigma_\pm$ and pressures
$p_\pm$ are not uniquely defined by the above relations, also for definite
values of ${\rm U}_\pm$.

Another quantity related with the counter-rotating motion is the specific
angular momentum of a particle rotating at a radius $r$, defined as
$h_\pm = g_{\varphi\varphi} U_\pm^\varphi$. We can write
\begin{equation}
h_\pm \ = \ \frac{{\cal R} e^{- \Phi} {\rm U}_\pm}{\sqrt{1 - {\rm
U}_\pm^2}} . \label{eq:moman}
\end{equation}
This quantity can be used to analyze the stability of the disks against
radial perturbations. The condition of stability,
\begin{equation}
\frac{d(h^2)}{dr} \ > \ 0 \ ,
\end{equation}
is an extension of Rayleigh criteria of stability of a fluid in rest in
a gravitational field; see, for instance \cite{FLU}.

Now we analyze the possibility of a complete determination of the
vectors $U_\pm^a$. As we can see, the constraint (\ref{eq:liga}) do
not determines ${\rm U}_\pm$ uniquely, and so there is a freedom in the
choice of $U_\pm^a$. The simplest, common, possibility is to take the
two counter-rotating tangential velocities as equal and opposite; that is,
\begin{equation}
{\rm U}_\pm \ = \ \pm \ {\rm U} \ ,
\end{equation}
so that (\ref{eq:liga}) is equivalent to
\begin{equation}
{\rm U}^2 \ = \ \left[ \frac{p_\varphi - p_r}{\sigma + p_r} \right] \ .
\label{eq:u2}
\end{equation}
This choice, commonly considered, leads so to a complete determination
of the velocity vectors $U_\pm^a$; however, this can be made only when
the above expression is positive definite. If it is not the case,
we will have a CRM valid only in a portion of the disk.

Another possibility, also commonly assumed, is to take the two
counter-rotating fluids as circulating along geodesics. Let be $\omega_\pm
= U_\pm^1/U_\pm^0$ the angular velocities obtained from the geodesic
equation for a test particle,
\begin{equation}
g_{11,r} \omega^2 + g_{00,r} = 0 , \label{eq:ecge}
\end{equation}
so that
\begin{equation}
\omega_\pm \ = \ \pm \ \omega \qquad , \qquad \omega^2 \ = \ - \
\frac{g_{00,r}}{g_{11,r}} \ .
\end{equation}
As the spacetime is static, the two geodesic fluids circulate with
equal and opposite velocities and so this is a particular case of the
above considered.

In order to see if the geodesic velocities agree with (\ref{eq:liga}),
we need to compute $f( {\rm U}_+ , {\rm U}_- )$. In terms of $\omega_\pm$
we get
\begin{equation}
{\rm U}_\pm \ = \ - \left[ \frac{W_1}{V_0} \right] \omega_\pm \ ,
\end{equation}
and so, using (\ref{eq:metdia}) and (\ref{eq:emtdia}), we can write
\begin{equation}
f( {\rm U}_+ , {\rm U}_- ) \ = \ \frac{ A + (\sigma + p_r - p_\varphi)
B}{ g_{11,r} {V_0}^2 } ,
\end{equation}
where
\begin{eqnarray}
A & = & g_{11,r} S_{00} + g_{00,r} S_{11} , \nonumber \\
	&	&	\nonumber	\\
B & = & g_{00} g_{11,r} + g_{00,r} g_{11} \nonumber .
\end{eqnarray}
Using the Einstein equations (\ref{eq:eins1}) - (\ref{eq:eins4}) and the
expressions (\ref{eq:emt1}) - (\ref{eq:emt3}) for the SEMT we can show that
\begin{equation}
f( {\rm U}_+ , {\rm U}_- ) \ = \ \left[ \frac{{\cal R}}{{\cal R}_{,r}
- {\cal R} \Phi_{,r}} \right] \frac{dp_r}{dr} \ ;
\end{equation}
that is, the counter-rotating fluids circulate along geodesics only
if the radial pressure is constant. In the general case, however, $f(
{\rm U}_+ , {\rm U}_- ) \neq 0$ for fluids circulating along geodesics
and so it is not possible to obtain a counter-rotating model with them.

As we can see of the above considerations, for disks built from generic
static axially symmetric metrics, the counter-rotating velocities are
not completely determined by the constraint (\ref{eq:liga}). Thus, the
CRM is in general undetermined since the energy density and isotropic
pressure can not be explicitly written without a knowledge of the
counter-rotating tangential velocities.

\section{A Family of Disks with Some Stable CRM}

We will now consider a simple specific example where we can obtain some
CRM with well defined counter-rotating velocities. In order to obtain
finite static disks with nonzero radial pressure we consider, following
reference \cite{GL1}, a solution of (\ref{eq:eins1}) - (\ref{eq:eins4})
obtained by taking
\begin{equation}
{\cal F} (\nu) \ = \ \nu + \alpha \sqrt{\nu^2 - 1} \ ,
\end{equation}
where $\alpha \geq 0$, so representing a thin disk located at $z = 0$,
$0 \leq r \leq 1$. Now we take a simple solution of Weyl equations
(\ref{eq:weyl1}) - (\ref{eq:weyl3}) given by \cite{RN}
\begin{subequations}\begin{eqnarray}
&&\Psi ({\cal R},{\cal Z}) \ = \ \frac{\mu}{2 k} \ \ln \left[ \frac{{\rm
R}_+ + {\rm R}_- - 2 k}{{\rm R}_+ + {\rm R}_- + 2 k} \right] \ , \\
&&	\nonumber	\\
&&\Pi ({\cal R},{\cal Z}) \ = \ \frac{\mu^2}{2 k^2} \ln
\left[ \frac{ ({\rm R}_+ + {\rm R}_-)^2 - 4 k^2}{4 \
 {\rm R}_+ \ {\rm R}_-} \right] \ ,
\end{eqnarray}\end{subequations}
where $\mu > 0$, $k = \sqrt{\alpha^2 - 1}$ and ${\rm R}_\pm^2 = {\cal
R}^2 + ( {\cal Z} \pm k )^2$.

From the above expressions, and using (\ref{eq:emt1}) - (\ref{eq:emt3}),
we can compute the energy density and azimuthal and radial pressures of
the disks. We obtain
\begin{subequations}\begin{eqnarray}
\sigma &=& p_r \left[ \frac{2\mu - \alpha}{\alpha} -  \frac{1 + \mu^2
r^2}{1 + k^2 r^2}\right] \ , \\
	&	&	\nonumber	\\
p_\varphi &=& p_r \left[ \frac{1 + \mu^2 r^2}{1 + k^2 r^2} \right] \ , \\
	&	&	\nonumber	\\
p_r &=& p_0 \left[ 1 + k^2 r^2 \right]^{(\mu^2 - k^2)/2k^2} \ ,
\end{eqnarray}\label{eq:depre}\end{subequations}
where $p_0 = 2 \alpha ( \alpha - k )^{\mu/k} \geq 0$ and $0 \leq r
\leq 1$.

We consider, in the first instance, two cases where we obtain simple
expressions for the SEMT. Let be $\alpha = 0$, so that $p_r = p_\varphi =
0$, and
\begin{equation}
\sigma \ = \ \frac{4 \mu e^{\mu\pi/2}}{(1 - r^2)^{(\mu^2 +
1)/2}} \ .
\end{equation}
We have a disk of dust with positive energy density, so in agree with
all the energy conditions \cite{HE}. On the other hand, the energy
density is singular at the edge of the disk. We can also see that the
constraint (\ref{eq:liga}) leads to ${\rm U}_\pm = 0$, so this is a ``true
static disk'', in the sense that we can not obtain a counter-rotating
interpretation for it. This case corresponds to the Bonnor and Sackfield
disk of reference \cite{BS}. For the second, simple, case we take $\mu =
k > 0$, so that $p_r = p_\varphi = 2\alpha(\alpha - k)$ and
\begin{equation}
\sigma = - 4(\alpha - k)^2 \ .
\end{equation}
The disks so are made of perfect fluids with constant energy density
and pressure. As $\sigma < 0$, the disks do not agree with the weak
energy condition, but $\varrho > 0$, as the strong energy condition
requires. For these disks we also have that ${\rm U}_\pm = 0$, so that
we do not have a CRM. These are also ``true static disks''.

For any other value of $\alpha > 0$ and $\mu \neq k$, the radial and
azimuthal pressures, $p_r$ and $p_\varphi$, are everywhere positive and
well behaved for all the values of $\mu$ and $\alpha$. For the effective
Newtonian density we obtain
\begin{equation}
\varrho \ = \ \frac{2 \mu p_r}{\alpha} \ ,
\end{equation}
so that is positive everywhere on the disk. Thus, the disks are
attractive, in agree with the strong energy condition. On the other
hand, is easy to see that $\sigma < 0$ when $r = 1$, for any value of
$\alpha$ and $\mu$, whereas that $\sigma > 0$ at $r = 0$ only if $\mu >
\alpha$. That is, in general, the energy density $\sigma$ is not positive
everywhere on the disk.

\begin{figure}
$$\begin{array}{cc}
\sigma \ : \ \mu = 1.5 & \sigma \ : \ \mu = 5.5	\\
\epsfig{width=2.5in,file=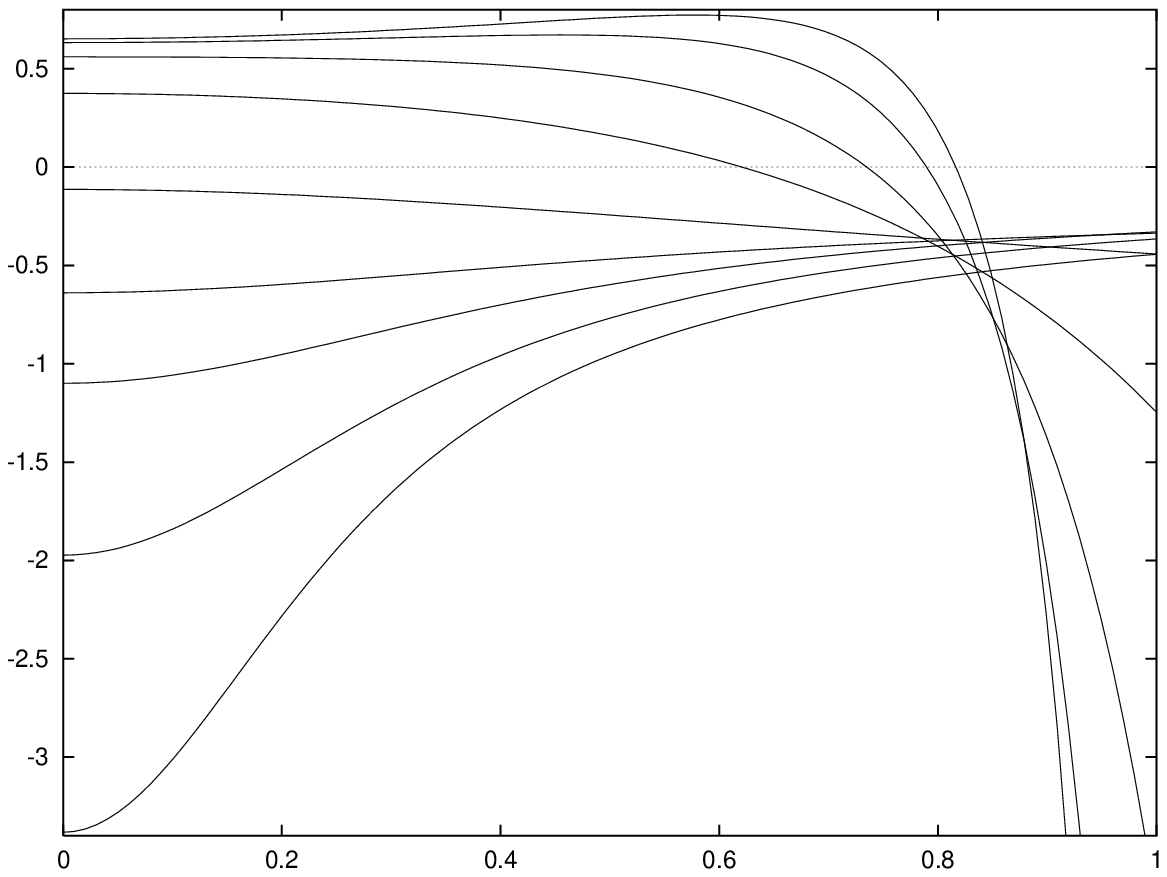}	&
\epsfig{width=2.5in,file=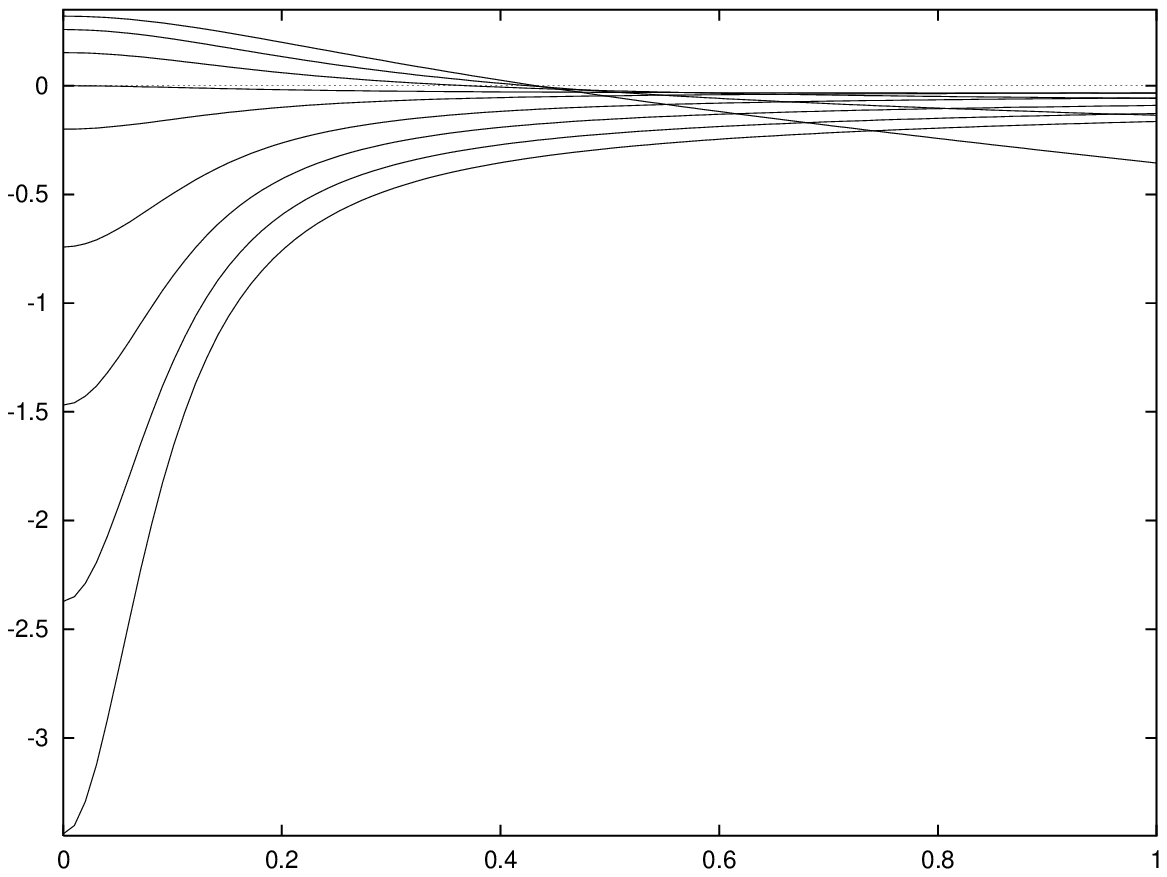}	\\
r & r
\end{array}$$
\caption{We first plot the energy density, $\sigma$, for a disk with
$\mu = 1.5$ and $\alpha = 0.5$, $0.6$, $0.8$, $1.1$, $1.6$, $2$, $2.3$,
$2.8$ and $3.5$. Then we plot $\sigma$ for a disk with $\mu = 5.5$
and the same values of $\alpha$.}\label{fig:densi}
\end{figure}

In order to study the behavior of the energy density and pressures we
perform a graphical analysis of them. We show, In Fig. \ref{fig:densi},
the energy density, $\sigma$, for disks with different values of $\mu$
and $\alpha$. We first plot $\sigma$ for a disk with $\mu = 1.5$
and $\alpha = 0.5$, $0.6$, $0.8$, $1.1$, $1.6$, $2$, $2.3$, $2.8$ and
$3.5$. Then we plot $\sigma$ for a disk with $\mu = 5.5$ and the same
values of $\alpha$. As we can see, for some values of $\mu$ and $\alpha$
the energy density is negative everywhere on the disks, whereas that for
other combination of the parameters the energy density is positive in
the central part of the disks, but negative in the edge. We also study
$\sigma$ for many other values of $\mu$ and $\alpha$ and, in all the
cases, we obtain a similar behavior.

\begin{figure}
$$\begin{array}{cc}
p_r \ : \ \mu = 1.5 & p_r \ : \ \mu = 5.5	\\
\epsfig{width=2.5in,file=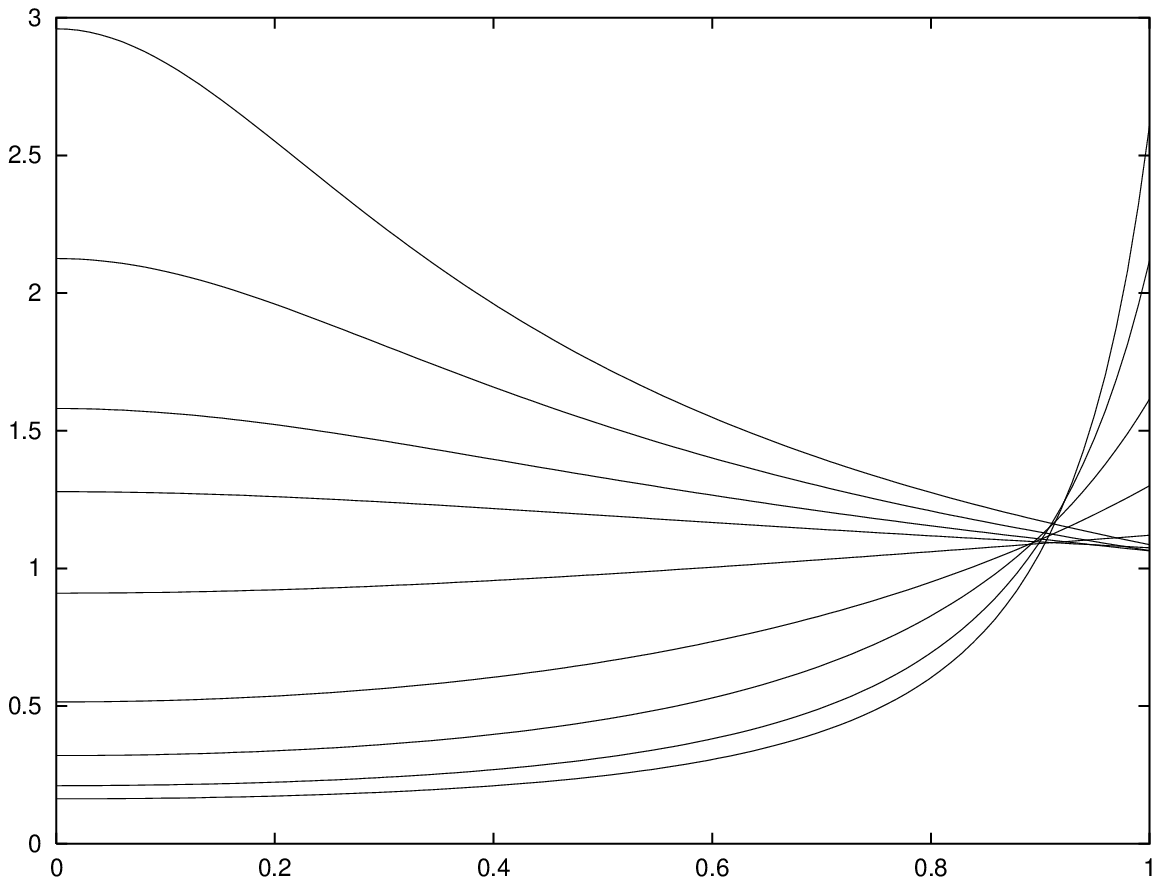}	&
\epsfig{width=2.5in,file=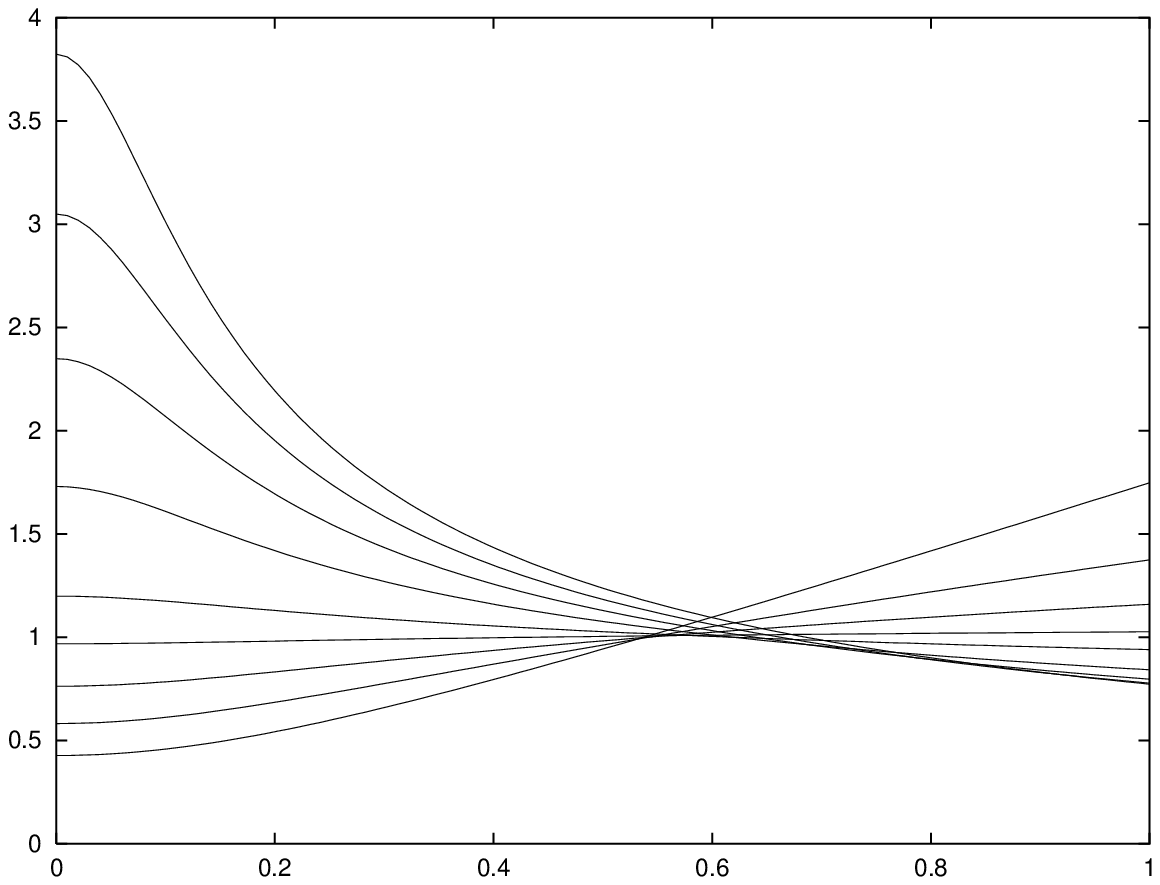}	\\
r & r
\end{array}$$
\caption{We first plot the radial pressure, $p_r$, for a disk with $\mu =
1.5$ and $\alpha = 0.5$, $0.6$, $0.8$, $1.1$, $1.6$, $2$, $2.3$, $2.8$
and $3.5$. Then we plot $p_r$ for a disk with $\mu = 5.5$ and the same
values of $\alpha$.}\label{fig:pradi}
\end{figure}

In Fig. \ref{fig:pradi} we depict the radial pressure, $p_r$, for disks
with different values of $\mu$ and $\alpha$. We first plot $p_r$ for a
disk with $\mu = 1.5$ and $\alpha = 0.5$, $0.6$, $0.8$, $1.1$, $1.6$,
$2$, $2.3$, $2.8$ and $3.5$. Then we plot $p_r$ for a disk with $\mu =
5.5$ and the same values of $\alpha$. As we said above, $p_r$ is positive
everywhere on the disks for all the values of $\mu$ and $\alpha$. Now,
for some values of the parameters $p_r$ have a maximum at $r = 0$ and
then decrease monotonly. On the other hand, with other values for $\mu$
and $\alpha$, $p_r$ is an increasing function of $r$. As with the energy
density, we also study the behavior of the radial pressure for many other
values of $\mu$ and $\alpha$ and all the cases considered presents the
same characteristics.

\begin{figure}
$$\begin{array}{cc}
p_\varphi \ : \ \mu = 1.5 & p_\varphi \ : \ \mu = 5.5	\\
\epsfig{width=2.5in,file=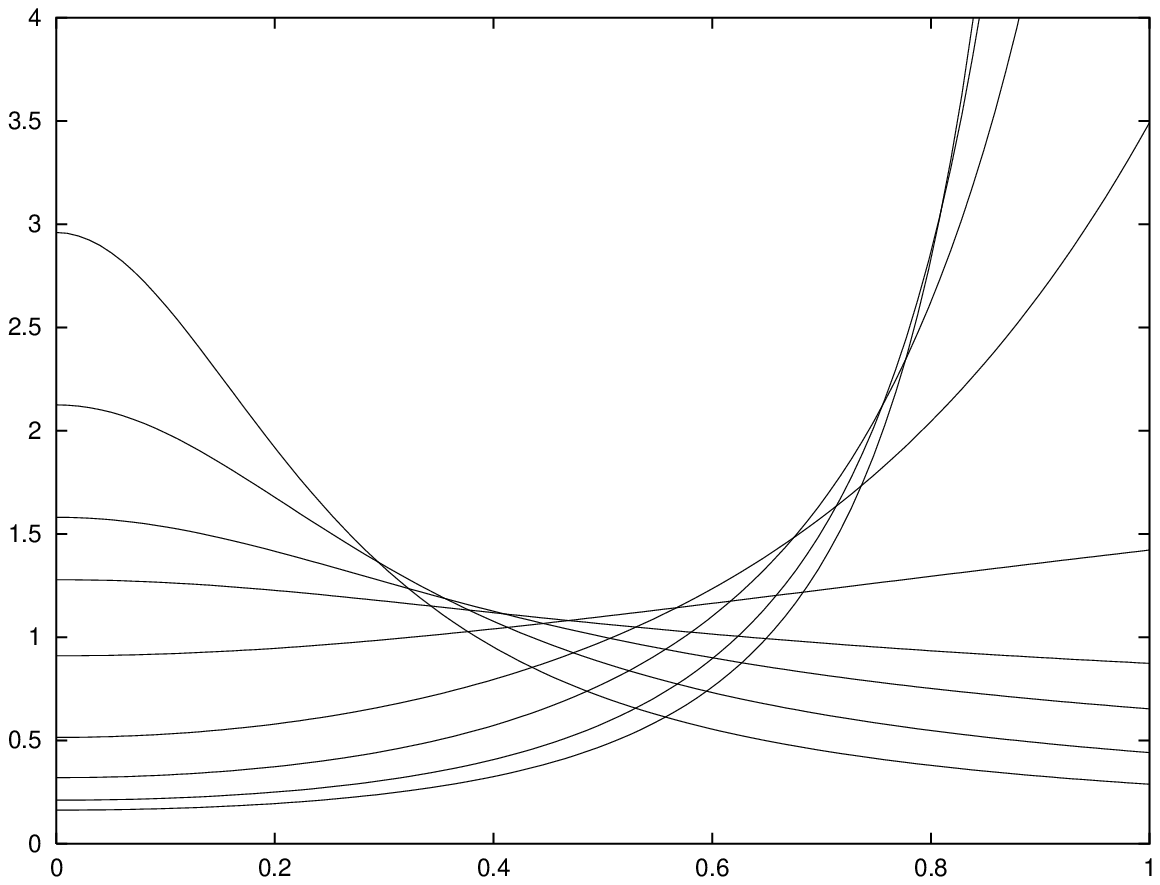}	&
\epsfig{width=2.5in,file=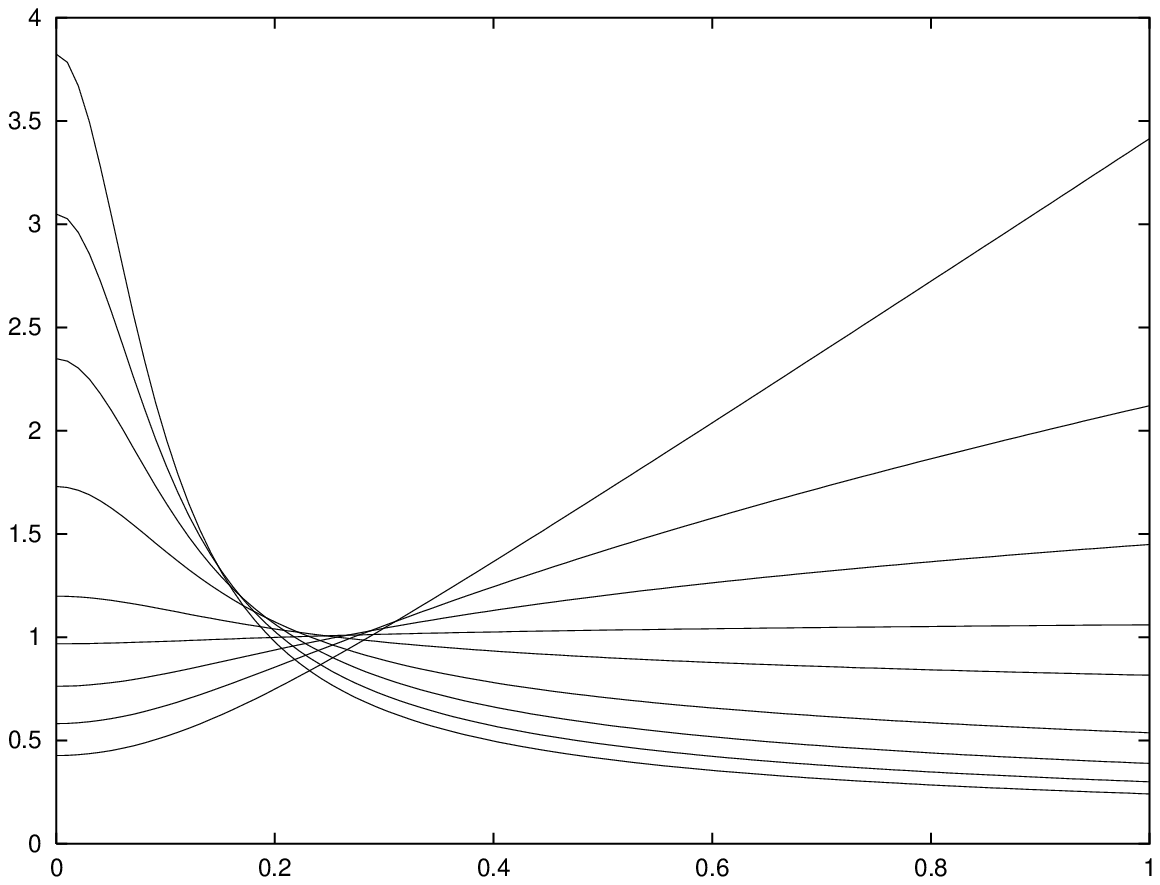}	\\
r & r
\end{array}$$
\caption{We first plot the azimuthal pressure, $p_\varphi$, for a disk
with $\mu = 1.5$ and $\alpha = 0.5$, $0.6$, $0.8$, $1.1$, $1.6$, $2$,
$2.3$, $2.8$ and $3.5$. Then we plot $p_\varphi$ for a disk with $\mu =
5.5$ and the same values of $\alpha$.}\label{fig:prazi}
\end{figure}

Now we study the behavior of the azimuthal pressure, $p_\varphi$,
in Fig. \ref{fig:prazi}. As in the previous analysis, we first plot
$p_\varphi$ for  disks with $\mu = 1.5$ and $\alpha = 0.5$, $0.6$, $0.8$,
$1.1$, $1.6$, $2$, $2.3$, $2.8$ and $3.5$. Then we plot $p_\varphi$
for disks with $\mu = 5.5$ and the same values of $\alpha$. We find
a behavior like with the radial pressure, with maximum at $r = 0$ for
some values of the parameters and increasing functions of $r$ for some
other values. We also consider many other values of $\mu$ and $\alpha$
and, as was the case with $p_r$ and $\sigma$, we obtain similar behavior.

\begin{figure}
$$\begin{array}{cc}
{\rm U}^2 & h^2	\\
\epsfig{width=2.5in,file=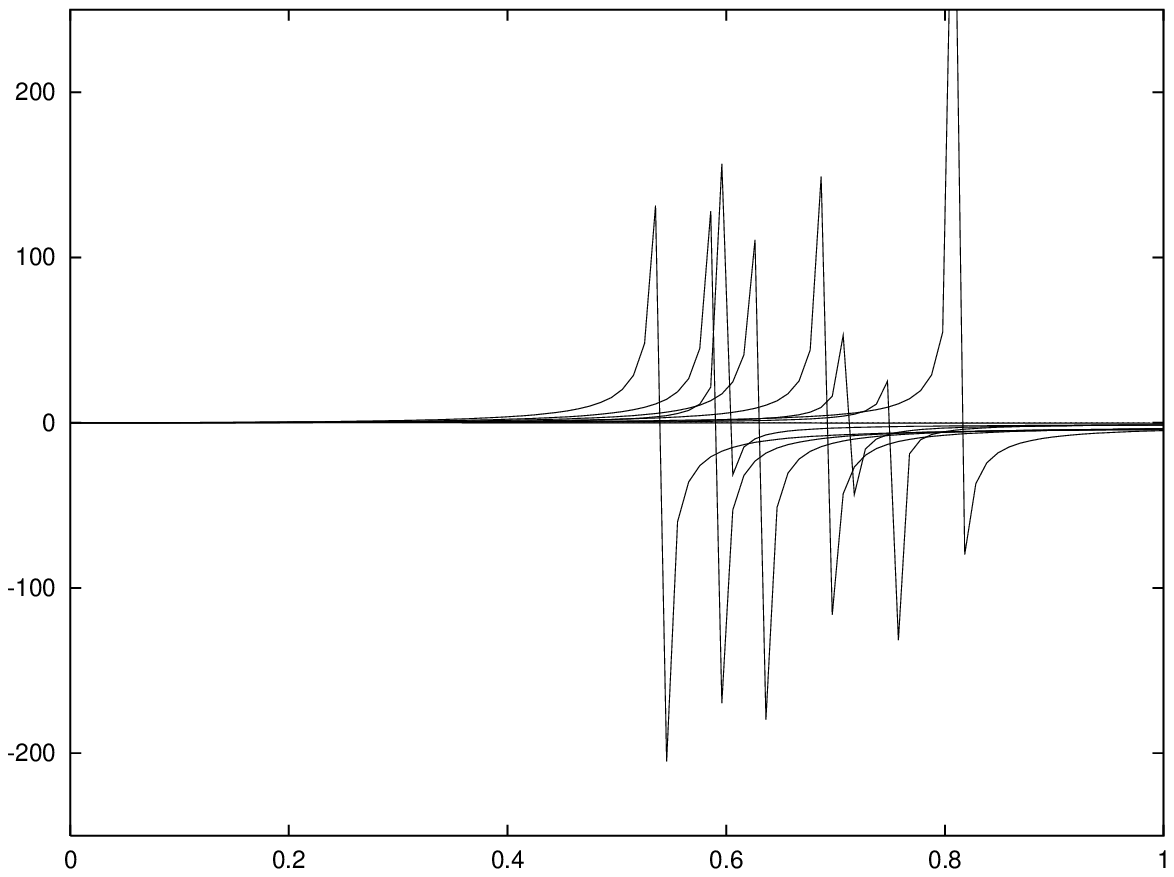}	&
\epsfig{width=2.5in,file=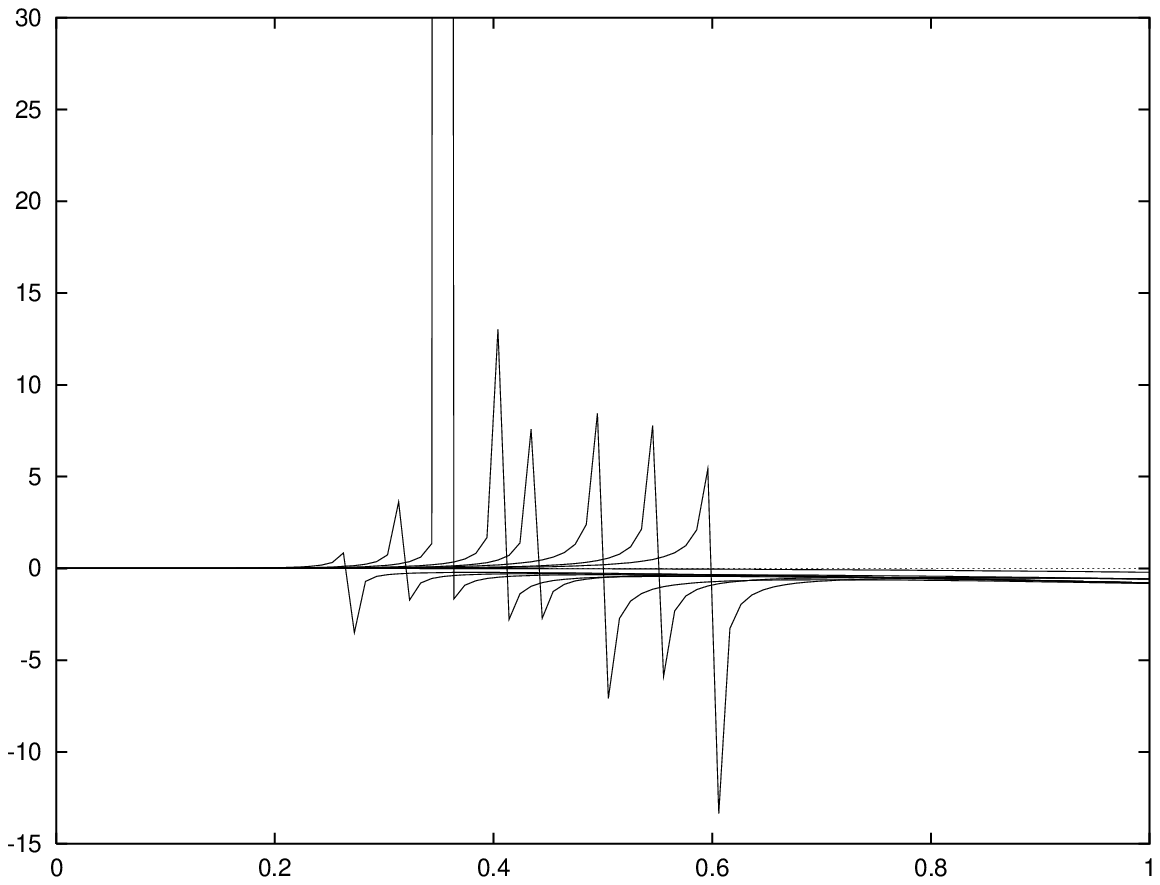}	\\
r & r
\end{array}$$
\caption{We plot ${\rm U}^2$ and $h^2$ for a disk with $\mu = 0.5$
and $\alpha = 0.5$, $0.6$, $0.8$, $1.1$, $1.4$, $1.7$, $2.2$, $2.3$
and $2.8$.}\label{fig:crm1}
\end{figure}

\begin{figure}
$$\begin{array}{cc}
{\rm U}^2 & h^2	\\
\epsfig{width=2.5in,file=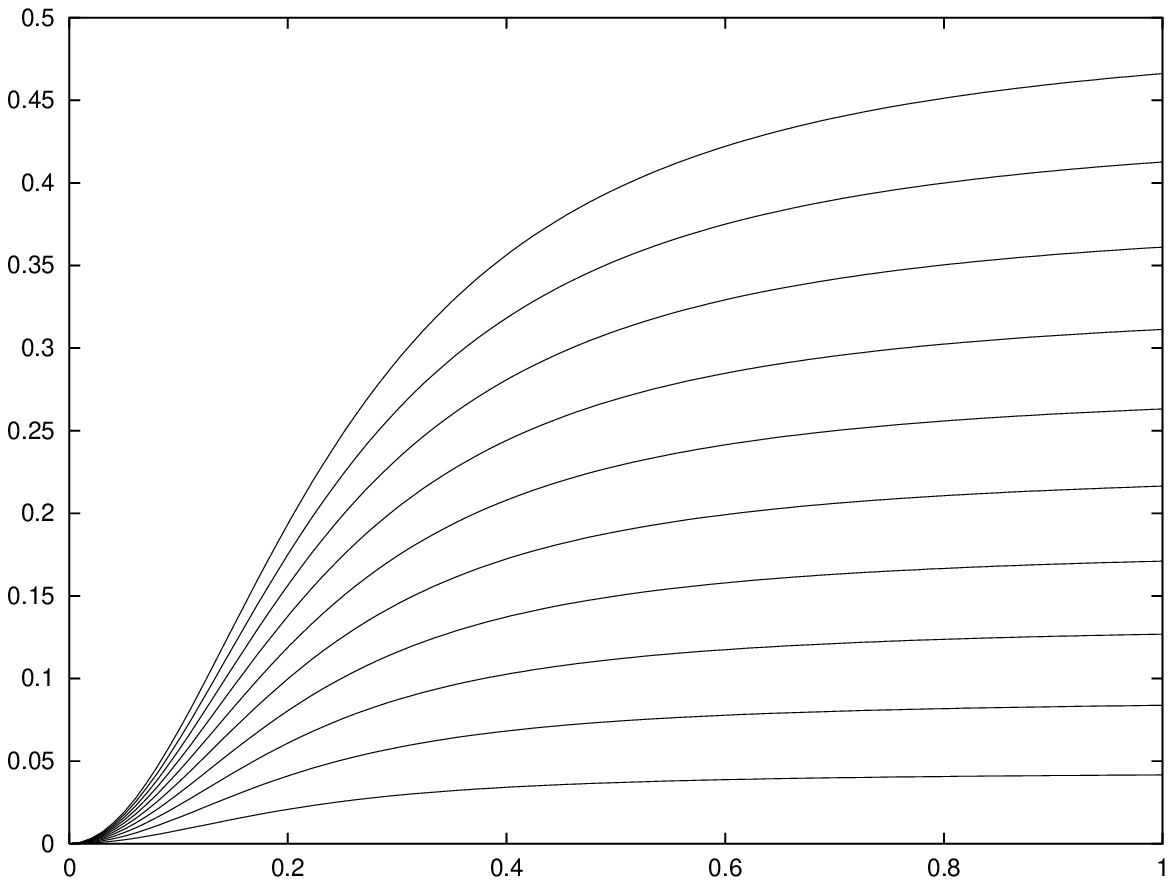}	&
\epsfig{width=2.5in,file=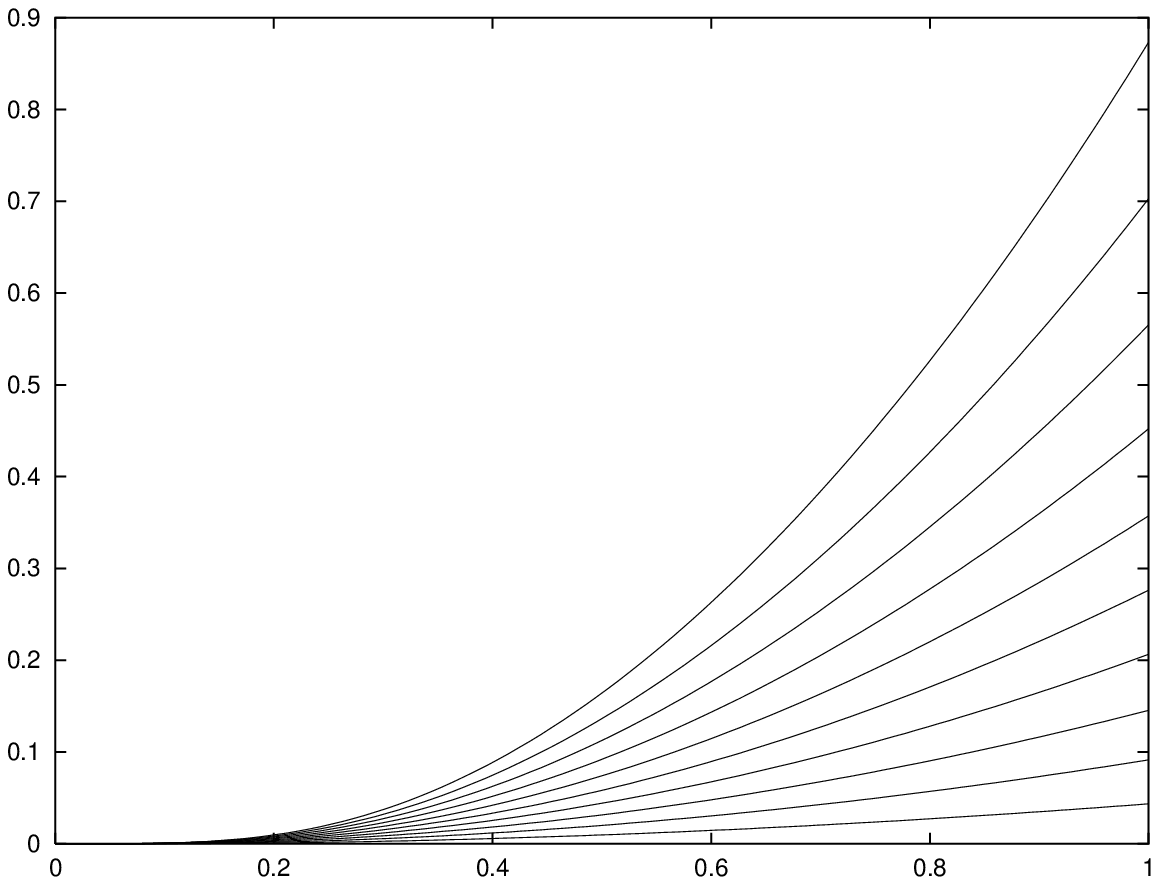}	\\
r & r
\end{array}$$
\caption{We plot ${\rm U}^2$ and $h^2$ for a disk with $\alpha = 5$
and $\mu = 5$, $5.1$, $5.2$, $5.3$, $5.4$, $5.5$, $5.6$, $5.7$, $5.8$
and $5.9$.}\label{fig:crm2}
\end{figure}

We now consider the CRM for the above disks. As $p_r$ is, in general,
dependent of $r$, we can not take the two counter-rotating fluids as
circulating along geodesics. However, we can test the possibility of
obtain a well defined CRM with equal and opposite velocities. In order
to do this, we can compute, from (\ref{eq:u2}) and (\ref{eq:depre}),
the tangential velocity ${\rm U}^2$ and obtain
\begin{equation}
{\rm U}^2 \ = \ \frac{\alpha (\mu^2 - k^2) r^2}{(2 \mu - \alpha) + \mu
(2 k^2 - \alpha \mu) r^2} \ ,
\end{equation}
and, in order to have a well behaved CRM, we impose the condition
\begin{equation}
0 \ \leq {\rm U}^2 \ \leq 1 \ .
\end{equation}
Again, is easier to do a graphical analysis, and so we study the
above relation for a lot of combinations of the parameters $\mu$ and
$\alpha$. In many of the cases we obtain functions with strong change in
the slope, with regions where ${\rm U}^2$ is negative and with ${\rm U}
> 1$. In order to see the kind of behavior that we have, we present a
sample plot in Fig. \ref{fig:crm1}, where we plot ${\rm U}^2$ for disks
with $\mu = 0.5$ and $\alpha = 0.5$, $0.6$, $0.8$, $1.1$, $1.4$, $1.7$,
$2.2$, $2.3$ and $2.8$. However, in some other cases we obtain ${\rm U}^2$
as well behaved functions of $r$, everywhere positive and increasing,
but always with ${\rm U}^2 < 1$. An example of these cases is shown in
Fig. \ref{fig:crm2}, where we plot ${\rm U}^2$ for disks with $\alpha =
5$ and $\mu = 5$, $5.1$, $5.2$, $5.3$, $5.4$, $5.5$, $5.6$, $5.7$, $5.8$
and $5.9$.

We can also compute the specific angular momentum of these CRM and,
using (\ref{eq:moman}) and (\ref{eq:depre}), obtain \begin{equation}
h^2 \ = \ \frac{\alpha (\alpha + k)^{\mu/k}(\mu^2 - k^2) r^4}{(2\mu -
\alpha) + [(2\mu - \alpha) k^2 - 2\alpha\mu^2] r^2} \ .  \end{equation}
Like with ${\rm U}^2$, the graphical analysis is better and, again,
we consider the above relation for many different values of $\mu$ and
$\alpha$. We also find, in many of the cases considered, strong changes
in the slope of $h^2$ as an indication of strong instabilities of the
CRM against radial perturbations. Also we found regions with negative
values of $h^2$, showing that the CRM can not be applied for these
values of the parameters. We shown an example of the cases considered
in Fig. \ref{fig:crm1}, where we plot $h^2$ for disks with $\mu = 0.5$
and $\alpha = 0.5$, $0.6$, $0.8$, $1.1$, $1.4$, $1.7$, $2.2$, $2.3$ and
$2.8$. We also obtain $h^2$ as increasing monotonic functions of $r$ for
some other values of $\mu$ and $\alpha$, so corresponding to stable CRM
for the disks. A sample of these cases is shown in Fig. \ref{fig:crm2},
when we plot $h^2$ for disks with $\alpha = 5$ and $\mu = 5$, $5.1$,
$5.2$, $5.3$, $5.4$, $5.5$, $5.6$, $5.7$, $5.8$ and $5.9$. These disks
are the same with well behaved tangential velocities, everywhere positive
and increasing functions of $r$.

\begin{figure}
$$\begin{array}{cc}
\sigma_+ + \sigma_- & \sigma_\pm + p_\pm	\\
\epsfig{width=2.5in,file=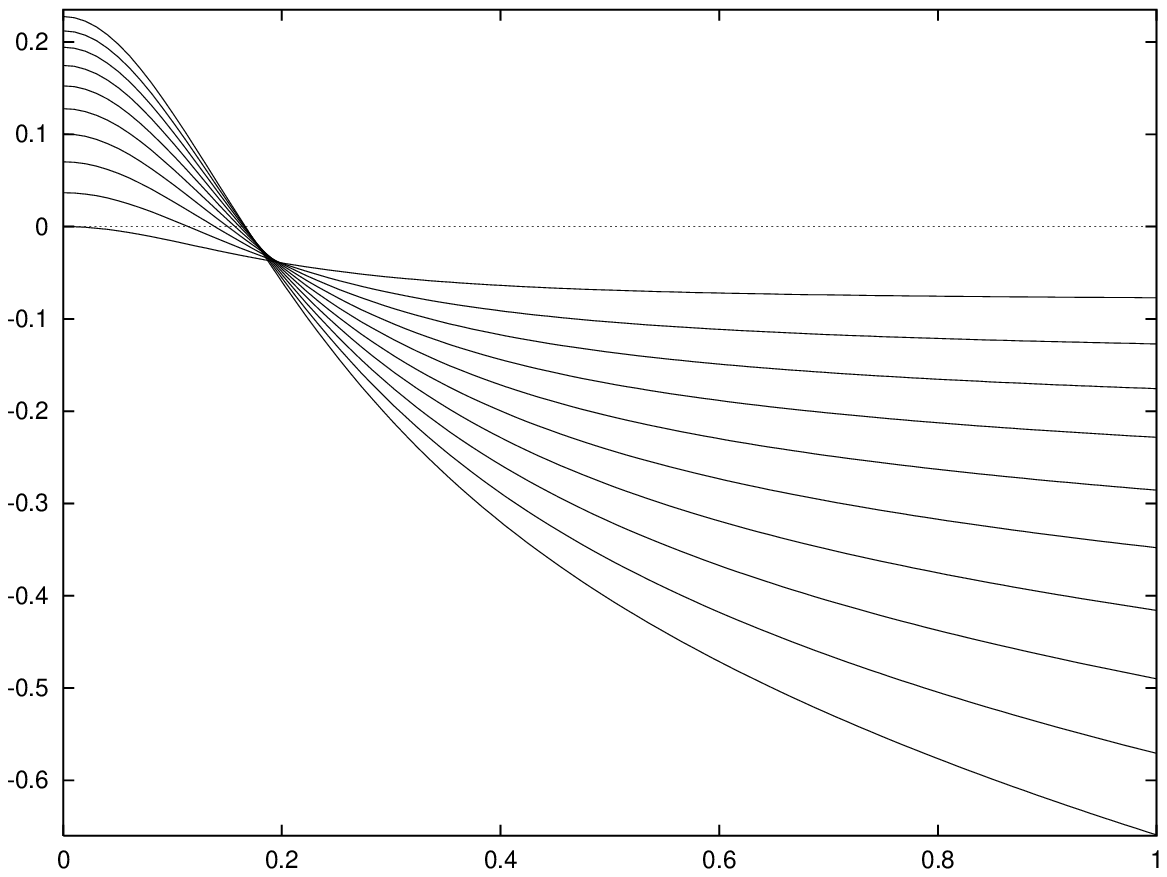}	&
\epsfig{width=2.5in,file=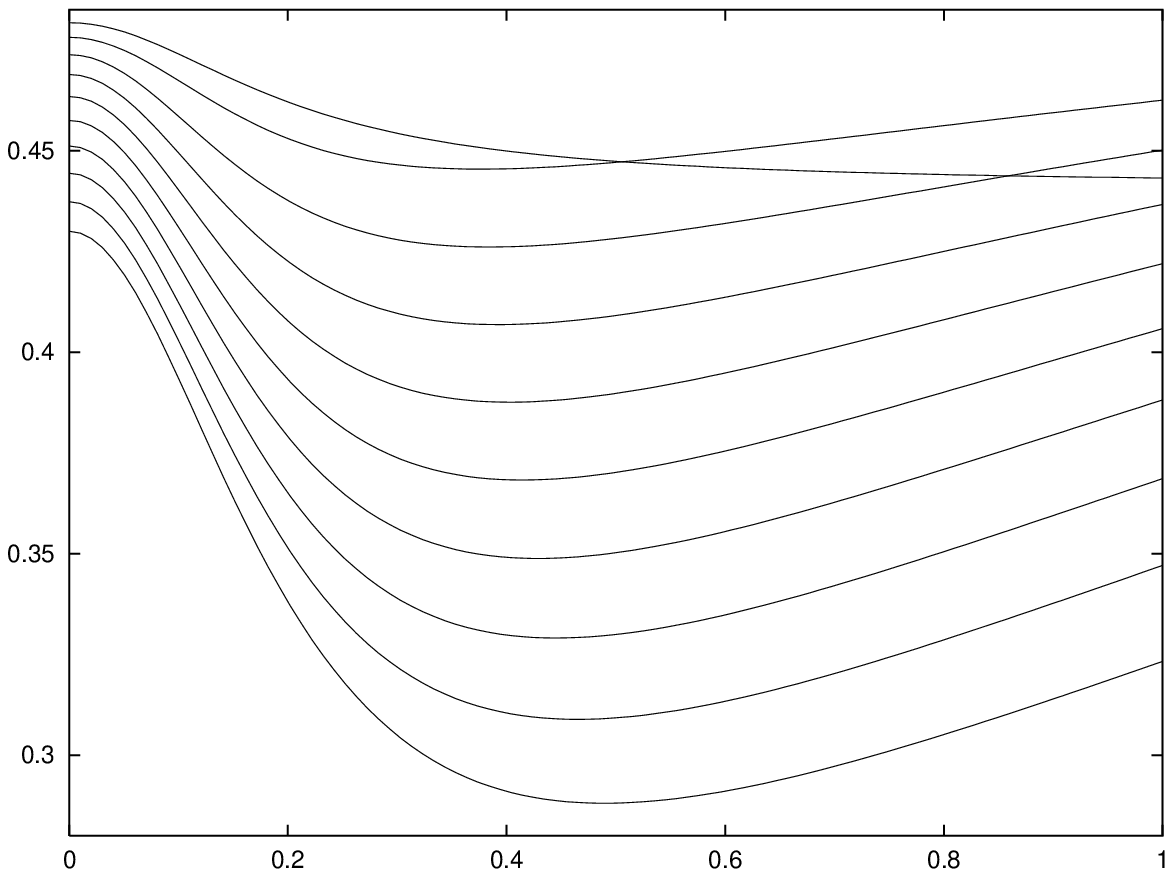}	\\
r & r
\end{array}$$
\caption{We plot $\sigma_+ + \sigma_-$ and $\sigma_\pm + p_\pm$ for a
disk with $\alpha = 5$ and $\mu = 5$, $5.1$, $5.2$, $5.3$, $5.4$, $5.5$,
$5.6$, $5.7$, $5.8$ and $5.9$.}\label{fig:dpcon}
\end{figure}

Finally, we can compute $\sigma_+ + \sigma_-$ and $\sigma_\pm + p_\pm$
for the above disks. Using (\ref{eq:dpcon}), (\ref{eq:u2}) and
(\ref{eq:depre}), we obtain
\begin{subequations}\begin{eqnarray}
\sigma_+ + \sigma_- &=& 2 \ p_r \left[ \frac{\mu}{\alpha} - \frac{1 + \mu^2
r^2}{1 + k^2 r^2} \right] \ , \\
	&	&	\nonumber	\\
\sigma_\pm + p_\pm &=& p_r \left[ \frac{2\mu + \alpha}{2\alpha} -
\frac{1 + \mu^2 r^2}{1 + k^2 r^2} \right] \ .
\end{eqnarray}\end{subequations}
We study the above relations for the values of the parameters $\mu$
and $\alpha$ that leads to well behaved tangential velocities and
specific angular momentum. As a sample, in Fig. \ref{fig:dpcon} we plot
$\sigma_+ + \sigma_-$ and $\sigma_\pm + p_\pm$ for a disk with $\alpha =
5$ and $\mu = 5$, $5.1$, $5.2$, $5.3$, $5.4$, $5.5$, $5.6$, $5.7$, $5.8$
and $5.9$. As we can see, for these values of the parameters, the total
energy density of the CRM, $\sigma_+ + \sigma_-$, is positive only in the
central region of the disks but negative in the rest having a maximum at
$r = 0$ and then decrease monotonly. On the other hand, $\sigma_\pm +
p_\pm$ is always positive for these cases. We see that the CRM obtained
are not only stable but in agree with the strong energy condition,
although not with the weak energy condition.

\section{Discussion}

We presented a detailed study of the Counter-Rotating Model for
generic finite static axially symmetric thin disks, with nonzero radial
pressure. A general constraint over the counter-rotating tangential
velocities was obtained, needed to cast the surface energy-momentum tensor
of the disk in such a way that can be interpreted as the superposition
of two counter-rotating perfect fluids. The constraint obtained is the
generalization of the obtained in \cite{GL2}, for disks without radial
pressure or heat flow, where we only consider counter-rotating fluids
circulating along geodesics. Also, we obtain expressions for the energy
density and pressure of the counter-rotating fluids in terms of the
energy density and azimuthal and radial pressures of the disk.

We shown that, in general, there is not possible to take the two
counter-rotating tangential velocities as equal and opposite neither take
the two counter-rotating fluids as circulating along geodesics. Thus,
for disks built from generic static axially symmetric metrics, the
counter-rotating velocities are not completely determined; that is, the
CRM is in general undetermined since the energy density and isotropic
pressure can not be explicitly written without a knowledge of the
counter-rotating tangential velocities.

An specific example was considered of a family of disks where we obtain
some stable CRM with well defined counter-rotating tangential velocities
and in agree with the strong energy condition, but with regions of the
disks where the energy density is negative, so in violation of the weak
energy condition. We also found some disks of the family presenting strong
changes in the slope of the tangential velocity and the specific angular
momentum, indicating so strong instabilities of the CRM, and with negative
${\rm U}^2$ and $h^2$ so that was not possible to obtain CRM for these
disks. Were found also two cases of ``true static disks'', in the sense
of ${\rm U}_\pm = 0$, and so there is not a possible CRM interpretation.
 
The generalization of the Counter-Rotating Model presented here to
the case of rotating thin disks with or without radial pressure is in
consideration. Also, the generalization for static and stationary disks
with magnetic or electric fields is being considered.

\subsection*{Acknowledgments}

O. A. Espitia want to thank a Fellowship from Vicerrectoria Acad\'emica,
Universidad Industrial de Santander.

\newpage

\end{document}